\documentclass[conference]{IEEEtran}
\IEEEoverridecommandlockouts
\pdfoutput=1
\usepackage{cite}
\usepackage{amsmath,amssymb,amsfonts}
\usepackage{algorithmic}
\usepackage{fancyhdr}
\usepackage{graphicx}
\usepackage{textcomp}
\usepackage{hyperref}
\usepackage{xcolor}
\def\BibTeX{{\rm B\kern-.05em{\sc i\kern-.025em b}\kern-.08em
    T\kern-.1667em\lower.7ex\hbox{E}\kern-.125emX}}

\begin{document}

This manuscript has been authored by UT-Battelle, LLC, under contract DE-AC05-00OR22725 with the US Department of Energy (DOE). The US government retains and the publisher, by accepting the article for publication, acknowledges that the US government retains a nonexclusive, paid-up, irrevocable, worldwide license to publish or reproduce the published form of this manuscript, or allow others to do so, for US government purposes. DOE will provide public access to these results of federally sponsored research in accordance with the DOE Public Access Plan (http://energy.gov/downloads/doe-public-access-plan).
\hfill \break \hfill \break \textbf{Submitted to the \textit{6th Annual Conference on Computational Science \& Computational Intelligence} held at Las Vegas, NV, USA on Dec 05-07 2019}

\title{DataFed: Towards Reproducible Research via Federated Data Management
}

\author{\IEEEauthorblockN{Dale Stansberry*, Suhas Somnath, Jessica Breet, Gregory Shutt, and Mallikarjun Shankar}
\IEEEauthorblockA{\textit{National Center for Computational Sciences} \\
\textit{Oak Ridge National Laboratory}\\
Oak Ridge, TN, USA \\
*Email: stansberrydv@ornl.gov}
}

\maketitle

\begin{abstract}
The increasingly collaborative, globalized nature of scientific research combined with the need to share data and the explosion in data volumes present an urgent need for a scientific data management system (SDMS).
An SDMS presents a logical and holistic view of data that greatly simplifies and empowers data organization, curation, searching, sharing, dissemination, etc.
We present DataFed - a lightweight, distributed SDMS that spans a federation of storage systems within a loosely-coupled network of scientific facilities.
Unlike existing SDMS offerings, DataFed uses high-performance and scalable user management and data transfer technologies that simplify deployment, maintenance, and expansion of DataFed.
DataFed provides web-based and command-line interfaces to manage data and integrate with complex scientific workflows.
DataFed represents a step towards reproducible scientific research by enabling reliable staging of the correct data at the desired environment.
\end{abstract}


\section{Introduction}
Several scientific domains are experiencing an explosion in the volume, variety, veracity and velocity of data owing to increased automation, increased computational power, and faster, higher resolution sensors and detectors in scientific instruments \cite{blair2014high,kalinin_bdsm}. 
At the same time, research is becoming ever more globalized, collaborative, and multidisciplinary, and there is an increasing need to publish the supporting datasets behind research findings \cite{nosek2015open_science}. Furthermore, scientific discovery using data analytics techniques like machine learning (ML) and artificial intelligence (AI) requires large volumes of high quality and well organized data. Prior research has shown that as much as 50-80\% of time is spent on data management and wrangling in most scientific research projects and this number is expected to rise \cite{marder_2015,furche2016data}. These factors are not only lowering scientific productivity but are also exacerbating the problem of poor reproducibility in science. The current state of the practice leads us to urgently seek a way to manage the lifecycle of data  with an effective Scientific Data Management System (SDMS) \cite{Moore:2000:DMS:647102.717555}, and use the SDMS as an essential component of the scientific process.


We require an SDMS to greatly alleviate common data handling and management challenges associated with large and/or complex corpora of data by providing an intuitive and holistic view of data and associated metadata supported by powerful but user-friendly organization, collaboration, and search/discovery capabilities. Data should be precisely identified and tracked to prevent accidental mishandling or misuse and to more easily support sharing with collaborators. Domain-specific metadata and provenance would need to be captured to provide additional context for data sets, and enable powerful AI-driven discoveries.

Many commercial and academic SDMS's are currently available and have been used in production-oriented and research laboratories respectively for many years \cite{quintero2013high,marini2018clowder, allan2012omero, arkin2016doe}. Commercial SDMS products typically support only a single organization or domain in a rigid manner and are oriented toward business processes rather than research.
Academic research focused SDMS solutions tend to be tied to specific scientific domains, are often challenging to deploy, and use non-scalable technologies that limit the pervasiveness and widespread acceptance of such services - resulting in disjoint data silos \cite{garonne2014rucio, rajasekar2007irods, marini2018clowder}.

In this paper we introduce DataFed - a general purpose SDMS tailored to scientific research with the overarching goals of eliminating data silos and enabling cross-organizational collaboration. DataFed was also designed to support the Findable Accessible, Interoperable, and Reusable (FAIR) data principles, which were proposed to facilitate open, collaborative, and reproducible scientific research \cite{wilkinson2016fair}. DataFed achieves these goals by providing both services and software technologies that help to bridge multiple scientific domains and organizational boundaries. By efficiently supporting scalable big data, and addressing the typical barriers to adoption such as complexity, rigidity, and administrative effort, DataFed aims to encourage cross-discipline and cross-domain data exchange. When deployed, DataFed forms a federation of geographically distributed research organizations, scientific facilities, and data repositories. The resulting data overlay establishes a scalable, multi-domain, scientific “data network” that greatly simplifies the data management aspects of cross-facility scientific research.

Our primary contributions are: (1) we describe the system architecture and implementation of an SDMS that offers a uniform (non-file-system-silo) view on data management across facilities, and (2) discuss specific benefits and use of DataFed to enable reproducible research over the data lifecycle. DataFed is available for use at \url{https://datafed.ornl.gov}. We proceed by first providing a high-level overview of DataFed in Section II, followed by the system architecture in Section III. Next, in Section IV, we discuss the benefits of using DataFed, and conclude with the future extensions planned for DataFed in Section V.


\section{System Overview}

DataFed is both a service and a software framework that integrates distributed scientific data repositories with centralized metadata and data management services to produce a federated, cross-organizational SDMS. DataFed is an “opt-in” data management network, where member organizations connect locally-managed data repositories to the federation, yet retain full control of the underlying storage systems (i.e. allocations and data policies). Central DataFed services are used to track, manage, and coordinate access to all data within the network.

\begin{figure}
    \centering
    \includegraphics[width=3in]{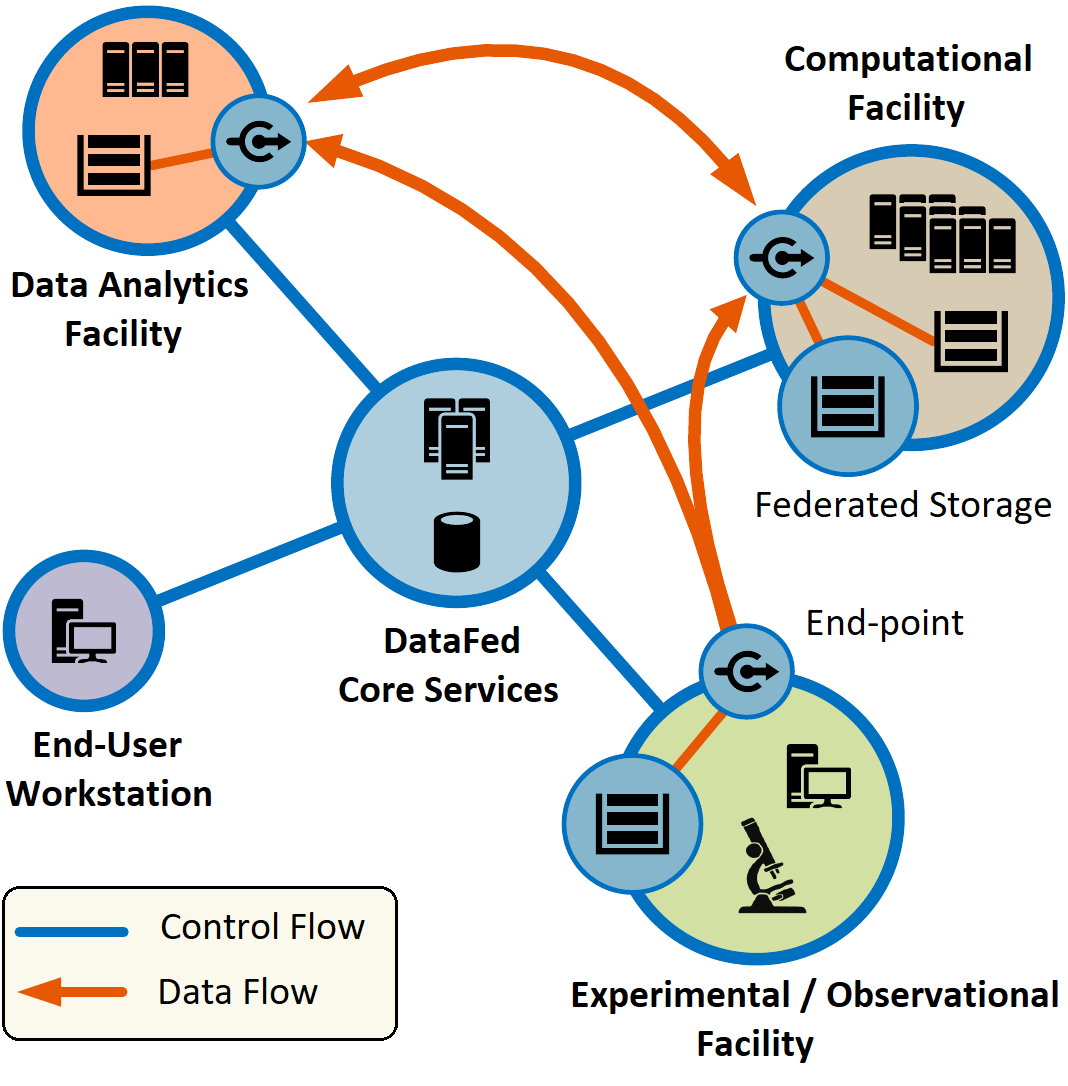}
    \caption{A conceptual overview of DataFed}
    \label{fig:high_level_overview}
\end{figure}

Figure~\ref{fig:high_level_overview} illustrates a high-level control-flow and data-flow view across different facilities using DataFed. Federating heterogeneous scientific facilities into a common data network enables the data access needs of complex cross-facility collaborations and workflows. Individual facilities may house one or more local data repositories, or may have none and rely on data in remote data repositories. Though member organizations may host various scientific resources such as experimental/observational instruments, or compute/analytics hardware, DataFed is only concerned with data access/management within and between these facilities - not with allocating the scientific resources themselves.

The Globus software suite~\cite{Allen:2017:GCS:3086567.3086570,chard2015globus} has become the de facto large-scale data communication mechanism of the scientific community by providing secure, performant, and managed data transfer services (based on  GridFTP \cite{allcock2003gridftp}) coupled with modern scalable user management technologies. By leveraging these popular Globus services, DataFed easily expands into existing Globus-enabled organizations and scales-out to support many users without creating the administrative bottlenecks that are common with older security models \cite{federated_identity}.

Unlike distributed file systems, DataFed {\em presents a logical view of data rather than a physical storage path to a named file}.
Internally, DataFed manages data as two distinct yet synchronized parts: a raw data object and a data record.
The raw data object can contain data in any format and is ingested from a source data file into a DataFed-managed data repository. The data record is used to uniquely identify the data, track administrative information, and store general- and domain-specific metadata.
DataFed's client interfaces or application programming interfaces (APIs) must always be used to access or modify managed data.

DataFed provides users with a variety of clients and interfaces - enabling simple and uniform data access from any connected resource within the DataFed federation, regardless of physical storage location or organization association. For example, a user can interactively identify and stage remote data to a local analytics resource for processing, or create scripts with DataFed APIs for batch processing and/or workflows on compute resources. DataFed can also be directly integrated into experimental facilities and/or data pipelines to automate data management. (We also employ a {\em Data Gateway} device that enables edge devices (scientific instruments) to submit data into the managed repositories of DataFed.)


\section{System Architecture}

\begin{figure*}[h]
  \centering
  \includegraphics[width=\textwidth,height=6cm]{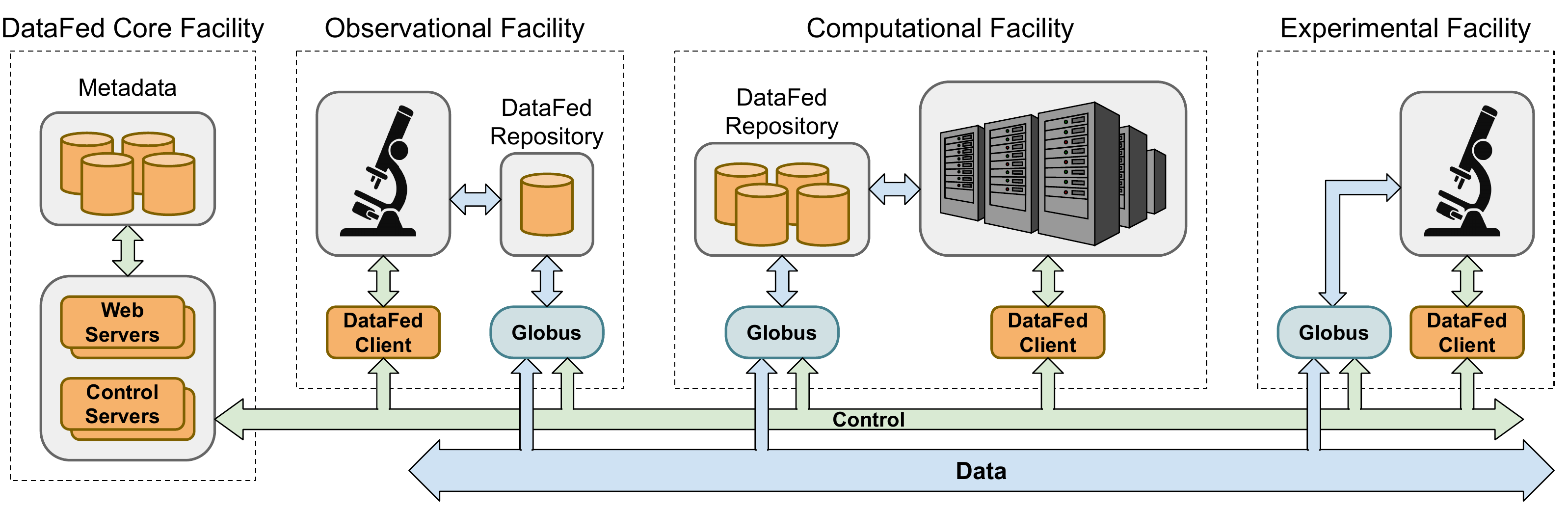}
  \caption{DataFed System Architecture}
  \label{fig:system_architecture}
\end{figure*}

The architecture of DataFed, shown in Figure \ref{fig:system_architecture}, consists of a ``Core Facility'' housing central DataFed services that manage and coordinate remote DataFed data repositories housed within member-organization-managed units, such as experimental, compute, and analytics facilities. To enable large data transfers DataFed requires that organizational facilities have a Globus endpoint (managed by the organization) connecting one or more local file systems to the DataFed network, and they must install a DataFed client for users to access and/or update remote data as illustrated for the ``Experimental Facility'' in Figure \ref{fig:system_architecture}. Optionally, a facility may house one or more local DataFed data repositories, which each require their own Globus endpoint (managed by DataFed) demonstrated by both the ``Observational Facility'' and ``Computational Facility'' in Figure \ref{fig:system_architecture}. DataFed services and clients communicate via a control protocol which is distinct from the GridFTP protocol. A user or process with sufficient permissions at one facility may access and/or update raw data housed within another facility's repository by requesting access through the Core services via a DataFed interface.

\subsection{Core Services}

DataFed core facility includes three distinct server types: 1) control servers, 2) web servers, and 3) a database server. Control servers are primarily responsible for orchestrating activities within the system that involve more that one component, such as data transfers between repositories or managing concurrent data access. Control servers also act as a gateway to microservices running within the underlying database. Web servers interface with control servers and are responsible for serving the DataFed Web Portal - which is a modern web application providing a graphical interface to all DataFed features. Finally, the database houses all of the records and metadata that describe the state of the entire DataFed network - including users, organizations, projects, repository configuration, metadata, access controls, and more. The raw binary data is stored in remote data repositories and not in the database. Given the criticality of the core database, the database should be deployed in a cluster configuration on highly resilient storage with a suitable data replication factor.

\subsection{Data Repository Servers}

DataFed data repository servers are a companion service that, along with a dedicated Globus endpoint, connects facility-local data storage to the DataFed network. The organization administering the repository may use any type or configuration of data storage system to support the repository so long as it is supported by Globus. The data repository software stack integrates with the associated Globus endpoint to route authorization requests to core services for all data access - enabling both fine-grained user authorization and concurrency controls.

\subsection{Client Interfaces}

Users can interact with DataFed via a command-line-interface (CLI) and a Web Portal. The CLI provides access to basic DataFed features for both interactive use and scripting via a Python 3 package that is available via the Python Package Index (PyPi) for all major operating systems. This python package provides both high- and low-level APIs that can be used to build custom applications that directly interface with DataFed.

The DataFed Web Portal provides users a simple yet powerful interface to manage their data through the use of modern web standards. Graphical elements, contextual menus, and drag-and-drop menus are used widely to simplify the interface and make operations intuitive. The focal-point of Web Portal is the Data Browser, which allows users to navigate through data, create and manage collections and projects, and share data with collaborators. Users can create data records, enter or edit metadata, view provenance graphs, transfer raw data, view publicly available data catalogs, and much more.

\subsection{Authentication and Data Transfer}

User account management based on certificates or encryption keys can become a significant administrative burden when managing a large number of users. The common approach of defining virtual organizations \cite{mowshowitz1997virtualorganizations} based on these authentication technologies can quickly become untenable due to the need to synchronize users across all member organizations. A more modern and scalable approach to user authentication is to use federated identity management \cite{federated_identity}, where users link multiple organizational accounts into a single composite identity. This approach enables mutual secure access between disjoint organizations despite differences in user accounting systems and security policies.

As mention previously, DataFed is based on Globus' federated user authentication and data transfer services, consequently benefiting from the availability, scalability, and performance advantages of Globus. As a consequence, DataFed users must also have a Globus account linked with one or more scientific institutes (universities, national laboratories, etc.). DataFed uses internal Globus endpoints for it's own data repositories and allows DataFed user to transfer data to/from external endpoints (either organizational or personal). DataFed manages and monitors all data transfers to or from DataFed data repositories while applying DataFed-specific concurrency controls and error handling policies.


\section{Benefits of DataFed for Scientific Research}

DataFed offers scientific researchers many of the benefits of production-oriented SDMSs such as unique data identification and tracking, abstraction of physical data storage, metadata and provenance capture, powerful organization and search capabilities, and data sharing with fine-grained data access controls. DataFed is also domain-agnostic, easy to learn and use, and scales-out across multiple organizations.

The logical representation of data as data records instead of files enables DataFed to offer a wide range of organizational possibilities beyond what is achievable through file systems alone, such as:

\begin{itemize}
    \item \textit{Collections} that offer a hierarchical, logical organization of data similar to file system directories, but are more flexible and descriptive. Collections can also be used to configure common access controls for contained records and support batch downloads.
    \item \textit{Dynamic Views} of data that can be created by users by saving metadata-based queries. Future versions will also allow tags to be used for dynamic views.
    \item \textit{Published Collections} organizes data that has been published by a user to one or more DataFed catalogs.
    \item \textit{Personal Allocations} that show data by repository to assist users with raw data storage monitoring and management.
\end{itemize}

When provenance information is captured, provenance graphs can be viewed from the web portal to allow users to dynamically explore related datasets, including those created by other users. Provenance information can be used to better understand complex interrelated data collections and serves as the basis for future DataFed data dissemination features.

\subsection{Query-able Structured Metadata}

DataFed data records support common top-level metadata schema consisting of title, description, keywords, and references (i.e. provenance) in addition to administrative fields such as identifier, owner, alias, etc.
DataFed does not impose schemas for the domain-specific structured metadata associated with data records. Users are free to define this metadata in any schema of their choosing and can use textual, numeric, Boolean, geospatial, and array value types.

The top-level metadata fields (textual values) are indexed by DataFed using a full-text search engine that supports root-word analysis.
Domain-specific metadata is not indexed by default but can be enabled by DataFed administrators after considering impacts on the query performance.
Despite not being indexed, arbitrary metadata query expressions may be used to search within domain-specific fields. For example, users can search a materials data collection to find datasets on samples of specific compositions tested over a specific temperature range. Users can limit searches to all personally-owned data, data owned by a specific project, data contained in a specific collection, etc.


\subsection{Collective Data Ownership with Projects}

DataFed offers collective ownership of data through the creation and use of DataFed Projects. These projects typically reflect real-world research being conducted by a team of researchers with one or more principal investigators (PIs). 
Projects are typically owned by a PI, who can add other DataFed users to the project and define project structure and access privileges for members.
Since projects have their own repository allocations, project data does not count against member's personal storage allocation(s). Furthermore, projects enable joint administration of data by PIs and project members.

\subsection{Scripting, Batch Processing, and Workflow Support}

The DataFed Python client package provides both high- and low-level APIs that support non-interactive use cases such as custom applications, batch processing, integration with computational workflows, and full data pipeline automation. Running any application that requires user authentication in a non-interactive scenario requires local security credentials to be available to the application. DataFed greatly simplifies the process of installing such credentials (issuing a single CLI command) compared to the common approach of manually generating and installing certificates or encryption key pairs.
DataFed APIs can be used to ingest data and metadata generated by scientific instruments, stage data in compute or analytics environments, ingest datasets resulting from processing steps in complex workflows, capture provenance based on input-output relationships in processing steps, etc.

\subsection{Simplified and Scalable System Administration}

As mentioned earlier, DataFed uses Globus's federated identity management, which has already been adopted by many research organizations. Therefore, DataFed does not add any additional user administrative burden to these organizations. At the same time, DataFed users benefit by only needing to sign-on once to Globus via their regular institutional credentials.

\subsection{Pre-publication Data Catalogs}

DataFed provides a pre-publication data catalog feature that allows users and projects to advertise publicly accessible collections of living data to the wider DataFed community. These catalogs are organized using a community-driven taxonomy which permits users to browse by topic (e.g. - scientific domain) and search data records within topics using keyword, phrase, and metadata expressions.

\section{Example Applications}

\begin{itemize}
    \item \textbf{Data Ingestion} - Data portals or pipelines are often used to capture raw data generated by scientific instruments and move it to a centralized data repository \cite{androulakis2010mytardis}. DataFed can be incorporated into such data pipelines to capture operational (user name, project id, etc.) and domain-specific (experiment description, sample identifier and description, instrument calibration, etc.) metadata that simplify organization, discovery, and sharing of data.
    \item \textbf{User Facility Administration} - The DataFed Projects construct enables staff at user facilities, which often support hundreds of visiting researchers a year, to easily manage the data for users while allowing those users to control their data once their research is concluded (for example - moving data back to a home institute data repository).
    \item \textbf{Data Analytics} - The association of rich domain-specific metadata with raw data in DataFed data records can significantly alleviate the data wrangling necessary to create collections of examples that can be used to train AI models. Users could rapidly develop, iterate over, and improve their models by searching for applicable data collections using search queries or tags in DataFed, placing the collection in their compute environment regardless of its physical location and applying the model to the data collection. ML and AI agents can be deployed to automatically  extract features, identify anomalies and embed such knowledge as metadata in DataFed data records.
    \item \textbf{Schema Support} - DataFed can serve as the global collaborative platform for scientific communities to come together and develop schemas for their domain.  DL and natural language processing and can be applied to facilitate mapping of metadata from one schema to another.
    \item \textbf{Provenance Capture in Workflows} - In addition to capturing the resulting data assets created from a series of data processing steps in a workflow, DataFed can also capture rich provenance information such as the relationships between the data created and consumed at each step, including the processing parameters, algorithms / software used for processing, etc. The presence of such transparent and rich provenance information would not only enable other users to repeat or replicate data processing but also provide additional confidence for downstream users to use such datasets, as well as peer reviewers of articles / datasets to vet such datasets.
    \item \textbf{Scientific Companion} - The scientific community has frequently expressed their desire for a digital companion that can assist researchers in planning and steering experiments. Such a companion could mine metadata, provenance, and raw data present in DataFed to avoid unnecessary repetition of experiments, show results from similar experiments, suggest unexplored parameter combinations, etc.
\end{itemize}


\section{Development Roadmap}

The following list highlights SDMS features that enhance scalability, reliability, and usability:
\begin{itemize}
    \item \textbf{High Availability and Reliability} - Services are required to be resilient, prevent data loss, recover from failures, and preserve data integrity. In DataFed, this effort is primarily focused on clustered core and database services.
    \item \textbf{Schema Support} -  DataFed will permit users / communities to define metadata schemas to both validate records at the time of creation and to support domain-specific search wizards that will generate graphical input forms based on a specified schema. This capability will facilitate the evolution of schemas and may potentially be followed by the ability to map fields between overlapping schemas.
    \item \textbf{Data Dissemination and Bi-directional Annotations} - Much like journal articles, datasets too can become outdated, or found to be incomplete or erroneous. In such cases, downstream users producing derived data products of such datasets need to be informed of upstream changes to the data record status. Conversely, a downstream user may discover an issue with an upstream data record and wish to notify the associated owners about the issue. DataFed modifications will allow users to "subscribe" to "data events" associated with records or collections. Bi-directional annotations will be added to DataFed to permit up- or down-stream users to attach informational, warning, and/or error notifications to records and collections.
    \item \textbf{Publication Support} - DataFed will allow data records to be represented using Digital Object Identifier (DOI) numbers with links to raw data curated in data repositories external to DataFed. DataFed interfaces could still be used to access such data records from within the DataFed network. Furthermore, DataFed would facilitate far more nuanced searches than typically possible in external data repositories, given the availability of rich domain-specific metadata. 
    \item \textbf{Data Caching and Replication} - DataFed will automatically cache transient copies of remote raw data within high-use facilities based on policies and/or user preferences to improve system efficiency. A data replication feature will also be added to allow users to create redundant, but synchronized, copies of raw data on separate data repositories for increased data safety.
    \item \textbf{Tags and labels} - The attachment of tags and labels to data records will offer additional organizational options such as dynamic views based on tags and faceted searches. Tagged datasets will also augment the collections of examples (with labeled ground truth) for training AI models.
    \item \textbf{Improved Batch Import} - The existing batch import feature will be enhanced to simplify the process of importing large preexisting data collections into DataFed.
    \item \textbf{Multimedia Attachments} - DataFed will allow small multimedia assets, such as thumbnail images to be attached to data records to enhance browsing of data catalogs.
    \item \textbf{Metrics} - Data records and collections will support community-driven metrics such as numbers of downloads or subscribers, ratings, etc. to highlight the quality of datasets.
\end{itemize}

\section{Conclusions}

We present DataFed - a general purpose and domain-agnostic SDMS aimed at significantly alleviating the burden of data management to improve scientific productivity, lower barriers to cross-institutional and collaborative research, and facilitate data-driven scientific discovery. DataFed provides users with a logical view of data that abstracts the physical data storage and facilitates the capture and enrichment of metadata (operational and domain-specific) associated with the raw data. Users of DataFed benefit from provenance capture, powerful organization and search capabilities, fine-grained access control for sharing data with collaborators, unique identifiers for data records, etc. 

DataFed has been designed in compliance with FAIR data principles for facilitating open, collaborative, and repeatable research. By using DataFed to manage both data and execution contexts captured within software containers \cite{kurtzer2017singularity},  
DataFed represents an important step towards reproducibility in open scientific research by enabling users to easily, correctly, repeatably, and reliably
work with datasets within appropriate compute or analytic contexts.

DataFed's use of popular and scalable user authentication and data transfer services overcomes limitations of prior SDMSs and  simplifies deployment, maintenance, and expansion of the DataFed federation. Besides lowering barriers to adoption and deployment, DataFed also provides a user-friendly Web Portal for user-friendly management of data as well as a command line interface for integration with complex workflows or to create custom applications. We are continually improving the resilience, scalability, performance, and capabilities of DataFed. Interested readers are welcome to use DataFed at \url{https://datafed.ornl.gov}.

\section*{Acknowledgment}

This research used resources of the Oak Ridge Leadership Computing Facility (OLCF) and of the Compute and Data Environment for Science (CADES) at the Oak Ridge National Laboratory, which is supported by the Office of Science of the U.S. Department of Energy under Contract No. DE-AC05-00OR22725.

\bibliographystyle{IEEEtran}
\bibliography{output.bbl}

\end{document}